\documentclass[journal]{IEEEtran}

\usepackage{cite}
\usepackage[pdftex]{graphicx}
\usepackage{soul}
\usepackage{amsfonts}
\usepackage{amsmath}
\usepackage{bm}
\usepackage{stmaryrd}

\usepackage{dsfont}

\begin{document}

\title{Conservation laws and quantum error correction: \\ towards a generalised matching decoder}

\author{Benjamin J. Brown
\thanks{B. J. Brown is with IBM Quantum, T. J. Watson Research Center, Yorktown Heights, New York 10598, USA and IBM Denmark, Pr\o vensvej 1, 2605 Br\o ndby, Denmark.}
}

\markboth{Perspective article submitted to IEEE bits}%
{Shell \MakeLowercase{\textit{et al.}}: Bare Demo of IEEEtran.cls for Journals}

\maketitle

\begin{abstract}
Decoding algorithms are essential to fault-tolerant quantum-computing architectures. In this perspective we explore decoding algorithms for the surface code; a prototypical quantum low-density parity-check code that underlies many of the leading efforts to demonstrate scalable quantum computing. Central to our discussion is the minimum-weight perfect-matching decoder. The decoder works by exploiting underlying structure that arises due to materialised symmetries among surface-code stabilizer elements. By concentrating on these symmetries, we begin to address the question of how a minimum-weight perfect-matching decoder might be generalised for other families of codes. We approach this question first by investigating examples of matching decoders for other codes. These include decoding algorithms that have been specialised to correct for noise models that demonstrate a particular structure or bias with respect to certain codes. In addition to this, we propose a systematic way of constructing a minimum-weight perfect-matching decoder for codes with certain characteristic properties. The properties we make use of are common among topological codes. We discuss the broader applicability of the proposal, and we suggest some questions we can address that may show us how to design a generalised matching decoder for arbitrary stabilizer codes.

\end{abstract}


\IEEEpeerreviewmaketitle

\section{Introduction}

A quantum-computing architecture capable of performing large-scale algorithms will require the careful design of many interconnected components~\cite{Fowler12surface}. In addition to well-controlled quantum hardware that can realise error-correcting codes that can detect commonly occuring errors~\cite{Shor96, AharonovBen-Or97}, the system will also depend on supporting classical software, a central component of which is a decoding algorithm. Where quantum error-correcting codes are designed to perform error-detecting check measurements to identify the signatures of an error, a decoder then takes the syndrome data from the check measurements to determine the error that may have occurred. All together, the underlying physical hardware, the choice of quantum error-correcting code, and the performance of the chosen decoding algorithm will determine the practicality of realising a scalable machine that can process quantum information.

The surface code~\cite{Kitaev03, Dennis02} underlies many of the leading approaches to produce fault-tolerant quantum computing hardware. Its high fault-tolerance threshold, together with its low-weight error-detection measurements that are local on two-dimensional array of qubits mean that the code is very practical for experimental realisation~\cite{Fowler12surface}. In this perspective we discuss progress towards realising the surface code, with a focus on the different types of decoding algorithms that have been proposed for its realisation. A variety of proposals have been presented to deliver a practical, high-performance decoder to enable the surface code to tolerate a relatively large rate of local noise. These proposals include adaptations of classical decoding algorithms such as belief propagation~\cite{DuclosCianci10, Panteleev2021degeneratequantum}, as well as new solutions to approximate maximum-likelihood decoding using tensor network methods~\cite{Bravyi14} and sampling using a Metropolis-Hastings algorithm~\cite{Wootton2012, Hutter14}.

Reviewing different methods of decoding is beneficial, not only to reflect on new ways to develop better decoders for the surface code, but they also provide us with a prototype decoder for more general low-density parity-check (LDPC) codes, of which the surface code is an archetypal example. See Ref.~\cite{Breuckmann21} for a recent review of progress on LDPC codes. Developing decoders for general LDPC codes may be invaluable in the future, as architectures based on these codes may be more hardware efficient than the surface code architectures we are currently developing to demonstrate scalable quantum computing. Resource efficiency will become an increasingly important issue as we scale our fault-tolerant quantum computing hardware to solve bigger and bigger problems.

We pay particular attention to the minimum-weight perfect-matching decoder~\cite{Edmonds65, Dennis02}. It is among the most well-studied decoders due to its practical implementation, and its demonstrated ability to tolerate a large rate of errors. Given its performance, it is important to understand the properties of the surface code that enables us to use this decoding strategy, as this can reveal how it may be generalised for use with other quantum error-correcting codes. The minimum-weight perfect-matching algorithm exploits fundamental structure of the surface code, namely, a parity conservation law among the defects of its error syndrome~\cite{Kitaev03, Brown20parallelized}. This conservation law means that, at a high level, error configurations can be identified by pairing syndrome defects. This problem is particularly well suited for matching algorithms~\cite{Dennis02}.

The parity conservation law of the surface code is intimately connected with a materialised symmetry~\cite{Kitaev03}, where the symmetry is expressed through a relationship among the stabilizers of the code. One of the main goals of this perspective is to begin to characterise symmetries more broadly for general stabilizer codes. We approach this in two ways. Firstly we look at generalisations of the minimum-weight perfect-matching decoder, including matching decoders for other codes~\cite{Wang10graphical, Delfosse14, Kubica19, Brown20parallelized, Nixon21, Sahay22}, together with bespoke matching decoders that are specialised to correct for noise models that demonstrate a specific structure or bias~\cite{Tuckett20, Bonilla-Ataides:2021}. Secondly, we propose a systematic method to construct a matching decoder for stabilizer codes with certain properties. We discuss the merits and applicability of the construction, and we try to highlight some questions that remain to be addressed to discover a general matching decoder.

The remainder of this perspective is structured as follows. In Sec.~\ref{Sec:Preliminary} we review the surface code together with the different noise models we often consider towards its development. In Sec.~\ref{Sec:Decoders} we review a number of different approaches that have been pursued to design and implement decoding algorithms for the surface code. We pay particular attention to the minimum-weight perfect-matching decoder. In Sec.~\ref{Sec:Symmetries} we discuss the underlying structure of the surface code, namely its symmetries, that permit the use of minimum-weight perfect-matching decoder, and we suggest ways of generalising this decoder to other families of codes. We consider generalisations of the minimum-weight perfect-matching decoder, first, by reviewing minimum-weight perfect-matching decoders that have been developed for other codes and, secondly, we propose a systematic way of constructing a matching decoder for codes with certain generic properties. We discuss the relative merits of this proposal, together with questions that need to be addressed to move towards a general construction for a matching decoder. Sec.~\ref{Sec:Discussion} offers some concluding remarks.

\section{Preliminaries}
\label{Sec:Preliminary}

Let us begin by reviewing the surface code, and discuss how it responds to different types of errors. We begin this section by introducing some basic notation needed to describe quantum error-correcting codes.

\subsection{Quantum states and Pauli matrices}

Quantum error-correcting codes, such as the surface code, encode quantum states over many qubits. See Ref.~\cite{NielsenAndChuang} for an introduction to quantum mechanics from the perspective of quantum information science. A qubit is a two-level quantum system whose pure state is represented by a unit vector in two-dimensional Hilbert space $ \mathbb{C}^2$. We define an orthonormal basis of vectors $|0\rangle  $ and $|1 \rangle$ to specify the state of a qubit in this complex vector space. A quantum state of $n$ qubits is specified in Hilbert space 
\begin{equation}
\mathcal{H}_n \cong \underbrace{\mathbb{C}^2 \otimes \mathbb{C}^2 \otimes \dots \otimes \mathbb{C}^2}_{n \textrm{ times}}.
\end{equation} 
We write quantum state vectors $|\psi\rangle \in \mathcal{H}_n$ such that $ |\psi\rangle = \sum_x c_x | x \rangle $ where we take the summaton over all bit strings $x = (x_1, x_2,\dots, x_n)$ of $n$ bits with bits $x_j \in \mathbb{Z}_2$ and we have complex numbers $c_x \in \mathbb{C}$ such that $\sum_x |c_x|^2 = 1$. We have also used the shorthand $|x\rangle = |x_1\rangle \otimes |x_2\rangle \otimes \dots\otimes |x_n\rangle  $.

Quantum states $|\psi\rangle$ are transformed with $2^n \times 2^n $ matrices, where unitary operators specify the evolution of quantum states and measurements are specified by projectors. It will be convenient here to specify matrices in terms of the Pauli group over $n$ qubits, $\mathcal{P}^{ \otimes n} = \mathcal{P} \otimes \mathcal{P} \otimes \dots \otimes \mathcal{P} $. The Pauli matrices $\mathcal{P} = \langle i, X,Z \rangle $ are a group of $2 \times 2$ matrices that are generated by the Pauli-X operator, $X$, the Pauli-Z operator, $Z$, and the complex phase, $i = \sqrt{-1}$. The canonical qubit basis states are eigenstates of the Pauli-Z operator, i.e., $Z |0\rangle =  |0\rangle$ and $Z |1\rangle =  -|1\rangle$, and the Pauli-X operator is such that $X |0\rangle = |1\rangle$ and $X |1\rangle = |0\rangle$. We also have the Pauli-Y operator such that $Y = i ZX$ and we have the identity operator $X^2 = Z^2 = \mathds{1}_2$. We write generating matrices of the the Pauli group $\mathcal{P}^{\otimes n}$ with the shorthand $X_j$ and $Z_j$ such that
\begin{equation}
X_j = \underbrace{\mathds{1}_2 \otimes \dots \otimes \mathds{1}_2}_{j-1 \textrm{ times}}  \otimes X \otimes \underbrace{\mathds{1}_2 \otimes \dots \otimes \mathds{1}_2}_{n-j \textrm{ times}},
\end{equation}
and
\begin{equation}
Z_j = \underbrace{\mathds{1}_2 \otimes \dots \otimes \mathds{1}_2}_{j-1 \textrm{ times}}  \otimes Z \otimes \underbrace{\mathds{1}_2 \otimes \dots \otimes \mathds{1}_2}_{n-j \textrm{ times}}. 
\end{equation}
The Pauli-X, -Y and -Z matrices are Hermitian and unitary operators with eigenvalues $\pm 1$.

Finally, it will be helpful to define the group commutator between two matrices. The group commutator between matrices $P$ and $Q$ is $[P,Q ] =  PQP^{-1}Q^{-1}$. For the case of Pauli matrices, the group commutator takes values $\pm 1$. We say that $P$ and $Q$ commute (anti-commute) if $[P, Q] = +1$ $(-1)$.

\subsection{The surface code}
\label{SubSec:SurfaceCode}

It is convenient to describe the surface code using the stabilizer formalism. A quantum error-correcting code protects a logical subspace, or code space $\mathcal{C}$, which is spanned by code states $|\psi\rangle$. Code states lie in the common $+1$ eigenvalue eigenspace of the elements of the stabilizer group $\mathcal{S} \subset \mathcal{P}^{ \otimes n} $; an Abelian subgroup of the $n$-qubit Pauli group such that $-\mathds{1} \not \in \mathcal{S}$. Specifically, we have the code space
\begin{equation}
\mathcal{S} = \left\{ S \in \mathcal{P}^{\otimes n} : S |\psi \rangle = (+1) |\psi \rangle \quad \forall |\psi\rangle \in \mathcal{C}  \right\}.
\end{equation}
The stabilizer formalism also specifies canonically anti-commuting pairs of logical Pauli operators $\overline{X}_j,\,\overline{Z}_j \in \mathcal{L} \subseteq \mathcal{P}^{\otimes n}$ with $1 \leq j \leq k$ such that $\left[ \overline{X}_j, \overline{Z}_l \right] = -1$ if $j = l $ and $\left[ \overline{X}_j, \overline{Z}_l \right] = +1$ otherwise, moreover, $\left[ \overline{X}_j, \overline{X}_l \right] = \left[ \overline{Z}_j, \overline{Z}_l \right] = +1$. These operators commute with all of the elements of $\mathcal{S} $ but are not themselves elements of $\mathcal{S}$, i.e., $\mathcal{L} = \mathcal{Z}(\mathcal{S})\backslash{S}$ where $\mathcal{Z}$ denotes the centraliser with respect to the Pauli group $\mathcal{P}^{\otimes n}$. Logical operators perform non-trivial rotations on states in the code subspace.

Quantum error-correcting codes are characterised by the set of parameters $ \llbracket n, k, d \rrbracket $ known as the code length, the code dimension and the code distance, respectively. The code length $n$ is the number of physical qubits used to produce the code, the code dimension $k$ is the number of logical qubits encoded by the code, and the code distance $d$ is the weight of the logical Pauli operator with the least weight, where the weight of an operator is the number of qubits on which it has non-trivial support, i.e., non-identity support.

\begin{figure}
\includegraphics{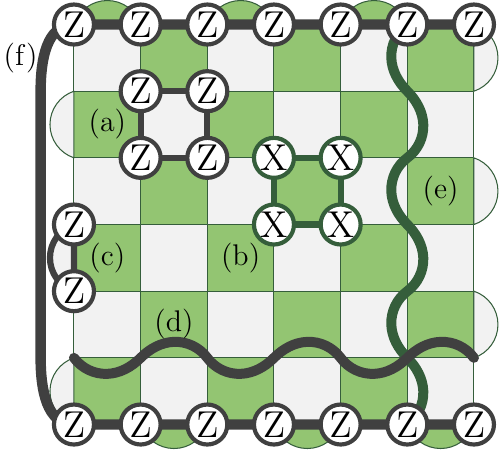}
\caption{\label{Fig:SurfaceCode} We define the surface code with a qubit on each of the vertices of a square lattice. The stabilizers of the surface code are defined with respect to the faces of the lattice, which are bicoloured. A Z-type~(An X-type) stabilizer is the product of Pauli-Z (Pauli-X) operators on the qubits at each of the corners of the light (dark) lattice faces, see (a) and (b), respectively. (c)~A weight-two boundary stabilizer on a Pauli-Z type boundary. (d)~and (e)~show the support of logical Pauli-Z and Pauli-X logical operators, respectively. They are such that the Pauli-X~(Pauli-Z) type logical operator terminates at a Pauli-X~(Pauli-Z) boundary. A boundary stabilizer~(f), $b = \prod_{f\in \text{grey}}S_f$, is supported on both of the Pauli-Z boundaries shown at the top and bottom of the figure.}
\end{figure}

The surface code is a Calderbank-Shor Steane (CSS)-type error-correcting code. This means that we can find a generating set of the stabilizer group such that all of its elements are the product of either Pauli-X operators, or the product of Pauli-Z operators. We will focus on a common implementation of the surface code, see Fig.~\ref{Fig:SurfaceCode} and its caption for a definition. The code has two types of boundaries that we will call X-type and Z-type boundaries. The two different boundary types are distinguished by the different types of stabilizers they support. Specifically, a Pauli-X (Pauli-Z)-type boundary supports weight-two Pauli-X (Pauli-Z) type stabilizers. The weight-two stabilizers are shown in the figure by rounded faces at the boundary, as shown at the top and bottom (left and right) lattice boundaries in Fig.~\ref{Fig:SurfaceCode}. We show a weight-two stabilizer explicitly in Fig.~\ref{Fig:SurfaceCode}(c).

The surface code we have defined encodes a single logical qubit. We define a pair of canonically anti-commuting Pauli-X and Pauli-Z logical operators, $\overline{X}$ and $\overline{Z}$, that are respectively the product of physical Pauli-X and Pauli-Z operators. Furthermore, they have a string-like support where the strings can terminate at lattice boundaries of appropriate type. For instance, a string of Pauli-X operators can terminate at the Pauli-X boundary, and a string of Pauli-Z operators can terminate at the Pauli-Z type boundary.  We show the support of a Pauli-Z (Pauli-X) type operator by the thick lines shown at Fig.~\ref{Fig:SurfaceCode}(d) (Fig.\ref{Fig:SurfaceCode}~(e)). It will be important later to note that there are many equivalent instantiations of the same logical operator, since a logical operator multiplied by a stabilizer group element will have the same action on the code space.

A quantum error-correcting code is designed to identify and correct for errors such that logical qubits remain well isolated from the physical environment. To determine the error that has occurred, we first measure stabilizer operators, and then use a decoding algorithm to interpret their outcomes and find an appropriate correction.
As the code space is defined as the common $+1 $ eigenvalue eigenspace of the stabilizer operators, a projection onto the $ -1 $ eigevnvalue eigenspace indicates an error has occurred. We say that face $f$ supports a defect if stabilizer $S_f$ gives the $-1$ measurement outcome. 
The list of stabilizer measurement outcomes is known as a syndrome.

Furthermore, it is important to note that stabilizer measurements project noise from more general quantum channels onto Pauli errors, as the eigenstates of the stabilizer measurements are of the form $E |\psi \rangle $ where $E$ is a Pauli operator and $|\psi \rangle$ is a code state. This allows us to concentrate our attention on Pauli errors, $E \in \mathcal{P}^{\otimes n}$, if we assume that more general noise channels have a very small support on the code space. Numerical simulations indicate that this assumption is well justified for surface codes with a large code distance~\cite{Darmawan2017, Bravyi:2018} undergoing local noise.

The CSS structure allows us to subdivide decoding into two classical problems, where the Pauli-X type stabilizers identify Pauli-Z errors, and Pauli-Z type stabilizers identify Pauli-X errors separately~\cite{Dennis02}. Pauli-Y type errors are regarded as a product of a Pauli-X error and Pauli-Z error under this separation. Using a duality between the two types of stabilizers, we will often discuss error correction with the surface code undergoing only Pauli-X type errors, sometimes known as bit-flip errors, but we note that an equivalent discussion will hold for Pauli-Z (dephasing) errors.

\subsection{The threshold theorem}

Scalable quantum computation has been demonstrated to be possible in the presence of errors due to the celebrated threshold theorem~\cite{Shor96, AharonovBen-Or97, Aliferis06}. It shows that the logical error rate can be made vanishingly small by increasing the distance of a quantum error-correcting code, provided the rate of error the physical qubits experience is below some threshold amount. To this end we require an efficient decoding algorithm to successfully recover encoded logical information. Given some prior information about the noise model, a decoder must interpret the syndrome information to find a correction operator $C \in \mathcal{P}^{\otimes n}$ such that $CE |\psi \rangle = |\psi \rangle$.

We find a geometric interpretation for the syndrome data of the surface code that we can exploit to design decoding algorithms~\cite{Dennis02}. With the exception of qubits at the Pauli-X boundaries of lattice, all single qubit bit-flip errors create a pair of defects of the lattice on the two adjacent Pauli-Z type stabilizers. This allows us to regard an error as a string, where the defects lie at the end points of the string. If we assume that errors are uncommon, this means the task of correcting errors far from the lattice boundary involves pairing nearby defects~\cite{Dennis02}. A bit-flip error at a Pauli-X type boundary of the lattice creates a single defect. This means single defects near to the boundary can be paired directly to the boundary.

\begin{figure}
\includegraphics[scale=0.5]{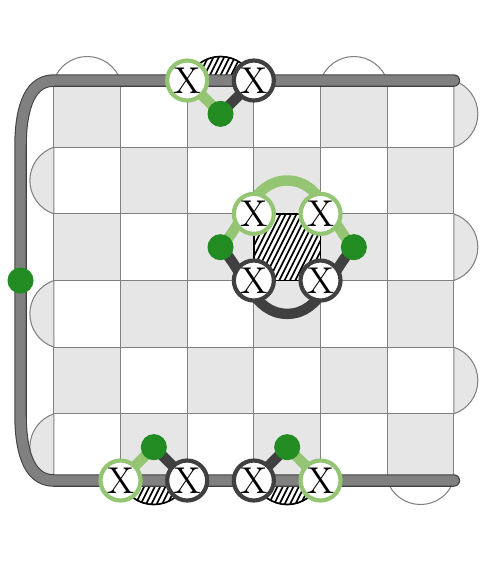}
\includegraphics[scale=0.5]{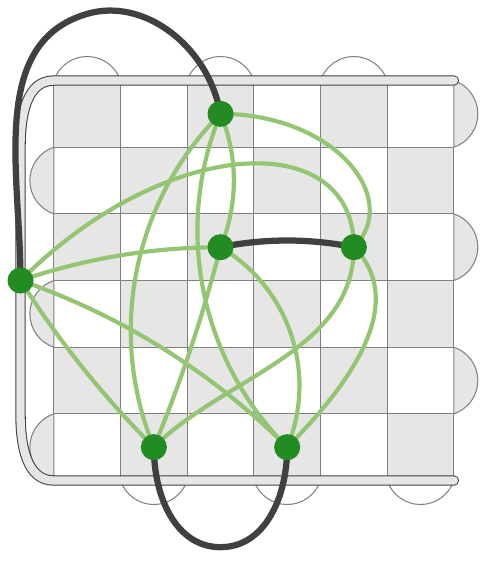}
\caption{Error correction with the surface code. (left)~An error is shown by gray Pauli-X operators, and the defects of its corresponding syndrome are shown by dark green spots associated to the faces of the Pauli-Z type stabilizers. Note that we have associated a defect to the boundary stabilizer $b = \prod_{f \in \textrm{grey}} S_f$ where we take the product over grey faces only. This is marked by the grey bar that covers the top and bottom boundary. A correction is shown by green Pauli-X operators. The product of the error and the correction is an element of the stabilizer group, i.e., the product of the hatched faces $S_f$. We therefore recover the encoded state successfully for this choice of correction. (right)~Decoding using the minimum-weight perfect-matching algorithm. The algorithm takes as input a graph with weighted edges, and it returns a perfect matching, i.e., a subgraph of the input graph where every vertex has exactly one incident edge. A minimum-weight perfect matching is such that the sum of the edges of the output subgraph is minimal. \label{Fig:ErrorCorrection}}
\end{figure}

In Fig.~\ref{Fig:ErrorCorrection}(left) we show an error with dark grey Pauli operators together with their corresponding syndrome shown in dark green. A correction, shown in green, is composed of strings that either connect pairs of defects, or connect single defects to the boundary. Both the error, and the correction operator, are underlaid with thick lines to emphasise the geometric structure of error-correction with the surface code. One can check that the error multiplied by the correction is an element of the stabilizer group. Specifically, we have that $CE = S \in \mathcal{S} $ where $S$ is the product of hatched face operators $S_f$. Decoding fails if the correction recovers the code space such that $CE \in \mathcal{L}$, see Ref.~\cite{Dennis02} and Appendix A of Ref.~\cite{Anwar14} for a detailed discussion on error correction with the surface code from the perspective of homology.

\subsection{Error models and unreliable measurements}

\label{SubSec:MeasurementErrors}

We consider a number of different error models to determine the performance of a quantum error-correcting code. First of all, we will consider an independent and identically distributed error model, where each qubit experiences a Pauli error with probability $p$. The simplest of which, that we have already mentioned, is the bit-flip error model, where a qubit experiences a Pauli-X error with probability $p$. We also consider the depolarising noise model where a qubit experiences a randomly chosen Pauli error with probability $p$. In general, an independent and identically distributed Pauli noise model will add a Pauli error $X_j$, $Y_j$ or $Z_j$ to qubit $j$ with probability $p_X$, $p_Y$ and $p_Z$, respectively, such that $p_X + p_Y + p_Z = p$. We say we have a biased noise model if $p_X$, $p_Y$ and $p_Z$ are not equal. For instance, physical architectures where the noise model is highly biased towards dephasing errors, e.g.~\cite{Darmawan21}, are such that $p_Z \gg p_X ,p_Y$. One could regard the bit-flip noise model as an infinitely biased noise model where $p_X = p$, and $p_Y = p_Z =0$.

\begin{figure}
\begin{center}
\includegraphics{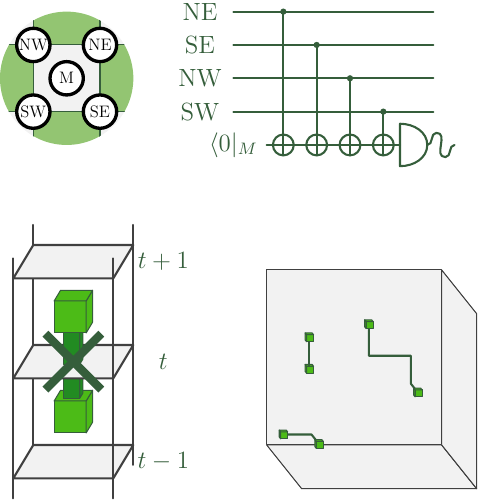}
\end{center}
\caption{Error correction with circuit-level noise. To measure a stabilizer $S_f$, an additional measurement qubit is added to each face $f$ of the surface code, labelled $M$ in the top-left part of the figure. The measurement qubit is initialised and entangled to the qubits at the corners of face $f$, and is subsequently read out to learn the value of a stabilizer operator. The circuit shown in the top right of the figure measures a Pauli-Z type stabilizer. To detect error events in the situation where measurement errors may occur, we measure each stabilizer repeatedly over time and compare consecutive outcomes. The bottom left figure shows a stabilizer measured three times at times $t-1$, $t$ and $t+1$, where each gray rectangle indicates a stabilizer reading is made. We suppose the stabilizer gives a false outcome at time $t$. Comparing the measurement outcomes at times $t-1$ and $t$, and then again at time $t$ and $t+1$ both reveal defects, marked by green cubes. In a $2+1$-dimensional space-time picture this measurement error can be regarded as a string that runs in parallel to the time direction. In general, to decode in the presence of measurement errors our task is to pair defects in a $2+1$-dimensional spacetime, such as that shown to the bottom-right of the figure. \label{Fig:FaultTolerant}}
\end{figure}

In addition to the code itself we also require additional apparatus to perform stabilizer measurements, and this apparatus may also experience errors. We show a stabilizer readout circuit in Fig.~\ref{Fig:FaultTolerant}. Errors that occur when we perform measurements can lead to a stabilizer measurement returning an incorrect outcome. This changes the way we approach error correction compared with the way we have discussed in the previous subsection. We must therefore design a procedure to account for measurement errors. Let us briefly discuss stabilizer measurement circuits and how we deal with unreliable measurements~\cite{Dennis02, Wang03}.

To identify a measurement error, we repeat the stabilizer measurement many times, and we report a defect if two consecutive readings of the same stabilizer disagree. Let us call the comparison between two adjacent stabilizer measurements a detection cell. We show two detection cells in Fig.~\ref{Fig:FaultTolerant}(bottom left). As we show, a measurement error creates a pair of defects at two consecutive times. In fact, we can regard the error as a string-like error where the string is aligned parallel with the time-like direction. In the presence of measurement errors, we recover the geometric picture for decoding discussed in the previous section, where events represented in a $2+1$-dimensional syndrome history are paired~\cite{Wang03}.

We should be mindful of other types of errors when we consider the circuits used for stabilizer readout. For instance, it is possible that an error can propagate through the readout circuit to introduce what is known as a hook error~\cite{Dennis02}. Stabilizer readout circuits and decoding algorithms have been designed to deal with the issue of hook errors~\cite{Fowler12surface}. These circuit-level details go beyond the scope of this perspective. We can often obtain compelling results using a simplified noise model that incorporates measurement errors, that we call the phenomenological noise model~\cite{Wang03}. This is where a measurement returns the incorrect value with probability $q = p$, where $p$ is the probability of a Pauli error occurring on a physical qubit per unit time.

\section{Decoding algorithms}

\label{Sec:Decoders}

Let us describe the decoding problem, and discuss the efficient algorithms that have been devised to correct errors the surface code experiences. We will pay particular attention to the minimum-weight perfect-matching decoder, as this motivates the discussion in the following section.

A decoder is given the syndrome $\sigma(E)$ for incident error $E$, together with prior information about the error model, and will then look for any correction $C$ with $\sigma(C) = \sigma(E)$ such that $CE$ acts trivially on the encoded state, i.e., $CE \in \mathcal{S}$. For an appropriate error model and assuming the rate of error is sufficiently low, the probability that the decoder successfully finds such a correction should rapidly approach unity as the code distance diverges in order for the code to demonstrate a threshold~\cite{Shor96, AharonovBen-Or97, Dennis02, Aliferis06}.

The errors that act on a quantum error-correcting code can be grouped into equivalence classes. Unlike classical error correction, there are many errors that can be introduced to a quantum error-correcting code that have an equivalent action. This is because errors are equivalent up to multiplication by a stabilizer operator. For instance, error $E \in \mathcal{P}^{\otimes n}$ has an equivalent action on the code as the error $ES$ for any $S \in \mathcal{S}$. Let us write $A \sim B$ if there exists a stabilizer $S \in \mathcal{S}$ such that $A = BS$, i.e., $A$ is in the same equivalence class of errors as $B$. We look for a correction $C$ such that $C\sim E$ since $CE = SE^2 = S \in \mathcal{S}$ assuming $C\sim E$ holds.

Different equivalence classes of errors differ by their action on the code space. Up to multiplication by stabilizers, the action of a correction $C$ changes compared to $E$ if and only if it differs from error $E$ by a logical operator, i.e., if $CL = ES$ for some choice of stabilizer $S \in \mathcal{S}$ such that $CE \in \mathcal{L}$. This distinction allows us to define different equivalence classes of errors $ \mathcal{K}_L $ with respect to some candidate choice of correction $C'$ that is fixed for all $ \mathcal{K}_L $, where each equivalence class is labeled by a logical Pauli operator $L \in \mathcal{L}$. We define
\begin{equation}
\mathcal{K}_L = \left\{ E \in \mathcal{P}^{\otimes n } : E \sim C'L  \right\}.
\end{equation}

For some correction $C'$, an optimal choice of decoder, otherwise known as a maximum-likelihood decoder, will consider the possibility of the occurrence of all errors belonging to $ \mathcal {K}_L$. Specifically, we look to maximise the value of $P_L$ for each logical operator $L$, such that
\begin{equation}
P_L = \sum_{E \in\mathcal{K}_L} \pi(E), \label{Eqn:ML}
\end{equation}
where we assume that we have prior knowledge to evaluate the probability $\pi(E)$ that the error $E$ was drawn from the distribution specified by the error model. We choose the correction $C = C' L$ for the maximal value of $P_L$

Given the exponential size of the calculation required to find the maximal value of $P_L$ as in Eqn.~(\ref{Eqn:ML}), in general it is inefficient to evaluate the values $P_L$ exactly. Indeed, it is important to find an efficient decoder in order to maintain the algorithmic speedup that is offered by a fault-tolerant quantum computer. In what follows we will discuss the different ways we have found to decode the surface code efficiently.

\subsection{The minimum-weight perfect-matching decoder}
\label{SubSec:MWPM}

We obtain an efficient high-performance decoder using the minimum-weight perfect-matching algorithm. Let us describe this decoder here. Our exposition may be unconventional compared to other descriptions of minimum-weight perfect-matching decoding in the literature, to help elucidate the generalisation for matching decoders we propose in the following section.

For reasonable error models such as local error models, the minimum-weight perfect-matching algorithm can determine an error that is highly probable to have occurred. In certain cases, for instance for the independent and identically distributed bit-flip noise model, it can obtain the maximal value of $\pi(E)$ over all $E$. It is known that this is sufficient for a decoder to demonstrate a threshold \cite{Dennis02}.

 A perfect matching is a subgraph of an input graph, such that every vertex of the output subgraph has exactly one incident edge, see Fig.~\ref{Fig:ErrorCorrection}(right). The minimum-weight perfect-matching algorithm~\cite{Edmonds65} takes as input a graph with weighted edges, and returns a matching of the input graph such that the sum of the weights of the edges of the output matching is minimal. See Ref.~\cite{Higgott21pyMatching} where an implementation of minimum-weight perfect-matching decoding algorithm is described.

We can adapt the minimum-weight perfect-matching algorithm to decode the surface code~\cite{Dennis02} by using the fact that error events introduce defects in pairs. Let us take the bit-flip error model as an example. It will be helpful re-express an error $E$ in terms of individual string-like error events $\alpha$ such that $ E = \prod_\alpha \alpha $ where each $\alpha$ creates exactly two defects for $\sigma(E)$.

For simplicity we assume that we can factorise the probability that error $E$ occurred as follows
\begin{equation}
\pi(E) = \prod_\alpha \pi(\alpha), \label{Eqn:MWPMassumption}
\end{equation}
such that we can evaluate the probability $\pi(\alpha)$ that the error model introduced error event $\alpha$ to the code. This assumption is justified for well-studied error models such as the bit-flip noise model, as well as phenomenological noise. This being said, a number of numerical analyses have demonstrated a threshold, even though they may not satisfy Eqn.~(\ref{Eqn:MWPMassumption}), see e.g.~\cite{Nickerson19}.

Given Eqn.~(\ref{Eqn:MWPMassumption}) holds, we can use the minimum-weight perfect-matching algorithm to obtain a maximal value for $\pi(E)$. To do so, we assign each defect a vertex of the graph that is input to the matching algorithm, we then produce a complete graph such that each edge connecting vertices $u$ and $v$ is assigned a weight $- \log \pi(\alpha)$, where $\alpha$ is a string-like error that introduced the events corresponding to vertices $u$ and $v$. The resulting minimum-weight perfect matching $M$ gives a solution with edges corresponding to string-like errors $\alpha \in M $ such that $  - \sum_M \log \pi(\alpha)$ is minimal or, equivalently, $\pi(E) = \prod_M \pi(\alpha)$ is maximal.

As an example, for the case of independent and identically distributed bit-flip noise for the surface code, it is common to choose edges with weights $ \sim \ell \log( (1-p ) / p)$~\cite{Dennis02, Wang03} where $p$ is the probability that a single bit-flip error event occurs and $\ell$ is the least number of errors that must occur such that an event $\alpha$ separates a pair of violated stabilizers by distance $\ell$. This is because the probability that a string of bit-flip errors occurring that separates a pair of violated stabilizers by a distance of $\ell$ is $\sim p^\ell / (1-p)^\ell $ for this noise model. It is worth noting that the performance of a decoder can be improved with refinements to edge weights, see e.g.~\cite{Stace10}.

To obtain a perfect matching, it is important that we perform matching over an even number of defects, such that we obtain an even number of vertices for the input graph. In the case of the surface code with boundaries, it may be that we have an odd number of defects associated to the lattice faces $S_f$. To deal with this, we introduce a boundary stabilizer
\begin{equation}
b = \prod_{f\in\text{grey}} S_f, \label{Eqn:BdryOp}
\end{equation}
where we take the product over all of the Pauli-Z type stabilizer generators $S_f$. The boundary operator is shown in Fig.~\ref{Fig:SurfaceCode}(f). This operator evaluates the parity of defects that have been measured by faces $S_f$. As such, it is not necessary to measure $b$ directly, as it can be inferred from the values of the face operators. In practice, this means that if we have an odd number of defects then we allow for one defect to be matched to a vertex associated to a Pauli-X boundary~\cite{Dennis02}. 

Let us formalise this statement. We have introduced a boundary operator $b$ to the set of Pauli-Z type operators $S_f$, such that we have a set of stabilizers $\Sigma$ whose product satisfies 
\begin{equation}
b \times \prod_{f\in\text{grey}} S_f = \mathds{1} . \label{Eqn:Symmetry}
\end{equation} 
It follows that the set of stabilizers $\Sigma $ that contains $b$ together with the grey face operators $S_f$ must detect an even number of defects, such that the product of their eigenvalues gives the outcome $+1$. Any other outcome will be inconsistent with the relationship given in Eqn.~(\ref{Eqn:Symmetry}). The set $\Sigma$ is an example of a materialised symmetry for the surface code~\cite{Kitaev03}.

The set of stabilizers associated to a materialised symmetry must give rise to an even number of defects. We call this a defect parity conservation law. As we have argued, a defect conservation law is a necessary condition to obtain a perfect matching for a decoding graph, as there needs to be an even number of vertices in the input graph to the perfect-matching algorithm to obtain a perfect matching.

It is common to use parity conservation laws to generalise the minimum-weight perfect matching algorithm. For instance, the detection cells defined in SubSec.~\ref{SubSec:MeasurementErrors} give rise to a parity conservation law among defects. This allows us to generalise the minimum-weight perfect matching decoder to the situation where measurements are unreliable~\cite{Wang03}, and even to deal with circuit level noise, see, e.g., Ref.~\cite{Fowler12surface}.

In addition to these works, the minimum-weight perfect matching decoder has been generalised to decode hyperbolic surface codes~\cite{Breuckmann_2017}, color codes~\cite{Wang10graphical, Delfosse14, Kubica19,Chamberland_2020, Beverland2021, Sahay22}, classical fractal codes~\cite{Nixon21} and more generally for fracton topological codes~\cite{Brown20parallelized}. It was proposed in Ref.~\cite{Brown20parallelized} that materialised symmetries may present a path to generalise the minimum-weight perfect-matching decoder for an arbitrary choice of topological stabilizer codes. In the next section we will discuss these results further to propose a construction for a minimum-weight perfect matching decoder for other codes.

\subsection{Fast decoding: clustering and the union-find decoder}

The minimum-weight perfect-matching algorithm gives rise to high-threshold error rates but, in practice, its polynomial time overhead may introduce a time latency that will lead to a bottleneck for high-speed quantum-computing architectures. As such, algorithms that are more efficient and simple have been proposed to overcome this problem.  A seminal result among these efforts is the union-find decoder~\cite{Delfosse2021}. The union-find decoder systematically groups defects into correctable clusters. In the case of the surface code, correctable clusters include an even number of defects. More generally, a correctable cluster should respect all of the materialised symmetries of an error-correcting code.

The union-find decoder runs in almost linear time in the code length due to the data structures used to identify clusters. Moreover, it demonstrates high thresholds that are comparable with those of the minimum-weight perfect-matching decoder~\cite{Delfosse2021}. It has also been demonstrated that the decoder can correct all Pauli errors of weight up to $d/2$~\cite{Delfosse2021}. The decoder has been generalised to the fault-tolerant case to account for varying probabilities of different types of errors using a weighted union-find algorithm~\cite{Shilin2020}.

\subsection{Approximating maximum-likelihood decoding}

Let us also discuss maximum-likelihood decoders. For some situations, efficient maximum-likelihood decoders are known for error correction with the surface code. Furthermore, we have also developed methods for obtaining the threshold obtained by a maximum-likelihood decoder, even for cases where no efficient maximum-likelihood decoder is known.

As discussed in the introduction to this section, a maximum-likelihood decoder must evaluate the probability of an exponentially large number of errors that are consistent with a given syndrome. However, we can obtain a good approximation for the performance of a maximum-likelihood decoder by sampling the most probable errors from a given equivalence class of errors. In Ref.~\cite{Dennis02} it was proposed that we can map maximum-likelihood decoding onto sampling from the partition function of a disordered classical statistical-mechanical model. Under this mapping, frustrated terms of a classical Hamiltonian correspond to a typical error configuration drawn from a given error model, and different spin configurations correspond to errors that are equivalent up to stabilizer operators. The Hamiltonian is sampled at the Nishimori temperature, e.g. $ \beta \sim -  \log p / (1-p) $ where bit-flip errors occur at rate $p$. See ~\cite{Chubb18} for a general formulation of the statistical mechanical mapping. Refs.~\cite{Hutter14} propose sampling common errors with a Metropolis algorithm to approximate maximum-likelihood decoding in a practical way.

Although the general problem of maximum-likelihood decoding is inefficient in general, some success has been made in obtaining efficient maximum-likelihood decoders. In Ref.~\cite{Bravyi14} two decoders are proposed; the first, an optimal decoder for the surface code undergoing bit-flip errors, and the second, a decoder suitable for more general error models that approximates maximum-likelihood decoding using efficient tensor network contraction techniques. While these decoders have been very important for benchmarking various codes, the nature of tensor network contraction algorithms means that it is difficult to generalise these decoders for higher-dimensional codes, or even the surface code undergoing measurement errors where we will typically consider decoding a $2+1$-dimensional syndrome. In future work it may be valuable to discover generalisations of these decoders for noise models that reflect all of the error processes that occur in the laboratory.

\subsection{Belief propagation}

Let us finally discuss decoding quantum codes using belief propagation; a standard approach for decoding classical codes in linear time~\cite{Gallager1962}. In such decoders, prior information about the error model is updated by passing messages in the form of probability distributions, that correspond to the local error model, between adjacent code variables and nodes of the syndrome. Nevertheless, the generalisation of belief-propagation decoders to quantum codes is non-trivial. This is due to the fact that quantum codes have many degenerate errors, i.e. equivalent errors that give rise to a common syndrome. This can lead message-passing algorithms to produce bad results, as they typically will fail to converge on a fixed error that is consistent with the syndrome of the code~\cite{Poulin08}.

A number of proposals have been presented to overcome the problem of degenerate error configurations using belief-propagation decoders for quantum codes. A recent proposal~\cite{Panteleev2021degeneratequantum} that has received considerable attention combines belief propagation with ordered-statistics decoding. Supplementing belief-propagation decoding with ordered-statistics decoding has a time complexity that is cubic in $n$~\cite{Panteleev2021degeneratequantum}. This decoder has demonstrated good thresholds for topological codes~\cite{Roffe2020decoding, Kuo2022}. Furthermore, the decoder serves as a versatile decoder that can compare the performance of LDPC codes more generally~\cite{Panteleev2021degeneratequantum}.

The renormalisation-group decoder is another approach to generalise belief-propagation decoding for quantum codes~\cite{DuclosCianci10}. This decoder uses message passing to coarse grain prior information about the error model to determine the error that has been introduced to a topological code. The use of coarse graining, or `renormalisation', means the decoder can be parallelised to run very quickly, with only a logarithmic time overhead in the code length. A renormalisation-group decoder has also been used to decode a phenomenological noise model in Ref.~\cite{DuclosCianci14}.

\section{Conservation laws and quantum error correction}
\label{Sec:Symmetries}

Numerous insights have been obtained about the surface code by regarding it as a phase of matter~\cite{Kitaev03}. Quantum phases are specified by a Hamiltonian $H$. This is an operator whose eigenvalues give the energy of eigenstates of $H$. For the surface code we write the Hamiltonian, $H = - \Delta  \sum_f S_f$, with constant $\Delta >0$ and where face operators of the surface code $S_f$ are defined in SubSec.~\ref{SubSec:SurfaceCode}. With this definition code states of the surface code $| \psi \rangle$ are ground states, or `least-energy' states, of the surface-code Hamiltonian. Other eigenstates of the surface-code stabilizers, $E |\psi \rangle $, are excited states of $H$. More specifically, we can regard faces with defects, where $S_f E |\psi \rangle= (-1) E |\psi \rangle$, as supporting quasiparticle excitations. These excitations are local quanta of energy with a well defined position, $f$, and energy, or `mass' $ 2 \Delta $.

The excitations of the surface-code Hamiltonian respect conservation laws whereby, in the bulk of the lattice, excitations can only be created in pairs~\cite{Kitaev03}. This is the same conservation law that we exploit to decode the surface code. Here we discuss how the matching decoder has been generalised by exploring the conservation laws for other codes.

A parity conservation law corresponds to a materialised symmetry~\cite{Kitaev03} among its stabilizer group. Let us propose a definition for a materialised symmetry. We say a materialised symmetry $\Sigma$ is a set of stabilizers $S \in \mathcal{S}$ that satisfies
\begin{equation}
\prod_\Sigma S = \mathds{1}. \label{Eqn:MatSymmetry}
\end{equation}

Eqn.~(\ref{Eqn:MatSymmetry}) implies a conservation law. Given a state $|\phi \rangle$ we have that $(\prod_\Sigma S)  |\phi \rangle = (+1)  |\phi \rangle$. It follows that the eigenvalues $s = \pm 1$ for all $ S \in \Sigma$ must multiply together to give $+1$; the eigenvalues of the identity matrix, $\mathds{1}$, as shown on the right-hand side of Eqn.~(\ref{Eqn:MatSymmetry}). We see then that Eqn.~(\ref{Eqn:MatSymmetry}) gives rise to a conservation law, because the stabilizers on the left-hand side must give rise to an even number of defects, i.e., stabilizer measurements that return the $-1$ eigenvalue outcome, in order for Eqn.~(\ref{Eqn:MatSymmetry}) to hold. We can therefore exploit symmetries to design a decoder that uses minimum-weight perfect matching.

A natural question to ask is how we might choose a materialised symmetry that is suitable to produce a high-performance decoder for a general choice of stabilizer code. There are a number of subtleties in the definition of a materialised symmetry proposed above, and the definition alone is not enough to produce a decoder in general. Certainly, one can find trivial examples of symmetries $\Sigma$ that will not give rise to practical decoders. In the remainder of this perspective we will discuss some observations that may be helpful to consider if we are to reach a general construction for a matching decoder. We will review different examples of decoders that make use of materialised symmetries, and we will conclude by proposing a construction that systematically reproduces the matching decoder for codes with certain properties. We discuss how these properties enable us to use the construction, and we present some questions we may need to address to find a systematic way of constructing a general minimum-weight perfect-matching decoder.

\subsection{Examples of matching decoders for other stabilizer codes} 

Let us discuss some examples where minimum-weight perfect-matching decoders have been proposed and implemented for other examples of stabilizer codes, in addition to the surface-code decoder discussed in detail in SubSec.~\ref{SubSec:MWPM}.

Several variations of minimum-weight perfect-matching decoders have been proposed for the color code~\cite{Bombin06}, see~\cite{Wang10graphical, Delfosse14, Kubica19,Chamberland_2020, Beverland2021, Sahay22} for examples of color-code minimum-weight perfect-matching decoders. See also Refs.~\cite{Vasmer21, Shutty22} where matching decoders for a hybrid color-surface code are proposed. The color code is a stabilizer code that is local in two dimensions. It gives rise to point-like defects on the faces of an appropriate lattice, where each defect has an associated colour according to the location of the face operator they occupy. In all of the examples of matching decoders, subsets of defects are matched depending on their colour. Fundamentally, the subsets of detection events that are matched correspond to some conservation law. See Ref.~\cite{Sahay22} where color-code decoding is discussed in the context of conservation laws and materialised symmetries explicitly.

It is also worth remarking that minimum-weight perfect-matching has been used to decode hyperbolic surface codes~\cite{Breuckmann_2017}. Importantly, these code families have a finite rate $R > 0$, which means that the code dimension $k$ scales linearly with the code length $n$ as $n$ diverges. This is an important indication that these methods for decoding are not limited to codes with only a small number of encoded logical qubits. 

\begin{figure}[b]
\begin{center}
\includegraphics{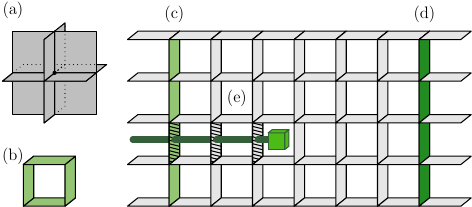}
\end{center}
\caption{\label{Fig:xCube} The X-cube model~\cite{Vijay16} can be defined with qubits on the cells of a cubic lattice Its stabilizers are the product of Pauli-X and Pauil-Z terms associated to vertices and faces of the lattice, e.g.~(a) and~(b). See Ref.~\cite{Brown20parallelized} for more details. A plane of the lattice is shown to the right of the figure. A Pauli-Z logical operator $\overline{Z}$ is supported on the light-green qubits below~(c). The product of the Pauli-Z stabilizers, such as that in~(b), associated to all of the complete cells in the figure give the product of the green faces below both (c) and (d). We call one such operator $b$ in the main text. The operator $b\overline{Z}$ is supported on the dark green qubits below~(d). We therefore think of the opertor $b$ as cleaning $\overline{Z}$ from the green qubits below (c) to the green qubits below (d). A Pauli-X error supported on the hatched qubits, (e), creates a single defect in the centre of the lattice. Pairing the single defect to the light-green qubits below (c) of boundary operator $b$, which must also be violated, indicates that an odd parity of bit-flips have occurred on the support of $\overline{Z}$, i.e, the green qubits below (c). We therefore obtain an estimate for the commutator of the error to the logical operator, thereby allowing us to find a correction that recovers the encoded state, as described in the main text.}
\end{figure}

More generally, several constructions have been proposed to realise two-dimensional stabilizer codes from other topological phases of matter~\cite{Kitaev03, Levin05, Kitaev06}. In this more general case, violated stabilizers that correspond to anyonic excitations of the underlying topological phase must respect certain fusion rules, see e.g., Appendix E of Ref.~\cite{Kitaev06}. These rules determine the allowed configurations of violated stabilizers. Naturally, these fusion rules that govern the allowed configurations of stabilizer violations can be interpreted as conservation laws. Decoding algorithms that correct more general anyon models have been proposed and implemented in Refs.~\cite{Brell2014, Wootton2014} where, essentially, all of these decoders look for a correction that respects the fusion rules that correspond to the anyon types of the violated stabilizers.

In addition to two-dimensional stabilizer codes that correspond to topologically ordered phases of matter, it is also interesting to explore more general codes. In Ref.~\cite{Brown20parallelized}, a minimum-weight perfect-matching decoder was proposed and implemented for a fracton topological code called the X-cube model~\cite{Vijay16}, see Fig.~\ref{Fig:xCube}. This is a three-dimensional model that is of interest at the level of condensed-matter physics as its quasiparticle excitations have constrained dynamics~\cite{Chamon05}. Specifically, quasiparticle excitations of the Hamiltonian of the X-cube model can only move in certain directions without increasing the energy of the system. These constrained dynamics have corresponding materialised symmetries that impose a parity conservation law for quasiparticle excitations among two-dimensional planes of the three-dimensional code.

We can regard the X-cube model as a stabilizer code by taking the commuting terms of its Hamiltonian as the generating elements of a stabilizer group describing a quantum error-correcting code. We can then use the symmetries of the X-cube model to design a matching decoder. In Ref.~\cite{Brown20parallelized}, a matching decoder is proposed that performs matching subroutines that respect the conservation laws of the two-dimensional materialised symmetries of the X-cube model. The decoder then takes the outputs from the different matching subroutines and combines them to obtain a correction. The results show that, by taking advantage of the additional structure that was imposed by the materialised symmetries of the X-cube model, the decoder gives rise to high threshold error rates.

\begin{figure}
\begin{center}
\includegraphics{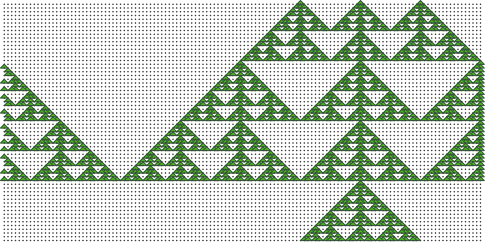}
\end{center}
\caption{\label{Fig:Fibonacci} The Fibonacci code is a classical code with geometrically local and translationally invariant weight-four stabilizers on a two-dimensional array of bits. Stabilizer generators are represented by green triangles. The local stabilizers of its symmetries can also be arranged as a fractal. We can exploit matching together with these symmetries to identify errors in individual bits. In Ref.~\cite{Nixon21} this observation was used to propose an iterative decoder that can tolerate very high-weight errors.}
\end{figure}

It will be interesting to discover matching decoders for more general stabilizer codes, such as other types of fracton topological codes~\cite{Vijay16}. For instance, Ref.~\cite{Haah11} proposes a so-called `type-II' fracton topological code, that has a distance that scales qualitatively better than the X-cube model, where logical operators are supported on a subset of qubits with a fractal structure. Ideally, we will find an efficient decoder that will correct $\sim d / 2$ errors, such that at low error rates, the logical error rate is minimal for local noise models. For high-distance codes such as a type-II fracton code, we should expect a very low logical failure rate. However, no such decoder has been implemented for a type-II fracton code.

In work towards a decoder for a type-II fracton topological code, Ref.~\cite{Nixon21} demonstrates a matching decoder for a two-dimensional classical fractal code. The decoder takes advantage of the fractal structure of the materialised symmetries of the classical code, see Fig.~\ref{Fig:Fibonacci}. In Ref.~\cite{Nixon21} a matching decoder was proposed and implemented that iteratively corrects individual bit-flip errors. Numerical results show that the logical failure rate of the iterative decoder decays rapidly in system size, to show that the decoder can take advantage of the high distance of the classical code. Given their similarities, these results demonstrate positive progress towards a high-performance decoder for a type-II fracton topological code.

\subsection{System symmetries}

The notion of a materialised symmetry is generalised to a system symmetry in Ref.~\cite{Tuckett20}. In this generalisation, we look for the symmetries of a code $\mathcal{S}$ that gives rise to an defect conservation law with respect to a specific error model $\mathcal{E} \subseteq \mathcal{P}^{\otimes n}$. The term system symmetry reflects that the materialised symmetries are described by the total system, that consists of both the stabilizer code $\mathcal{S}$ and the Pauli error model $\mathcal{E}$. This concept has proven valuable for finding quantum error-correcting codes that are specialised to correct for highly-biased noise models~\cite{Tuckett20, Bonilla-Ataides:2021, Srivastava2022xyzhexagonal, Dua2022, Xu22, SanMiguel22, Huang2022}.

\begin{figure}
\includegraphics{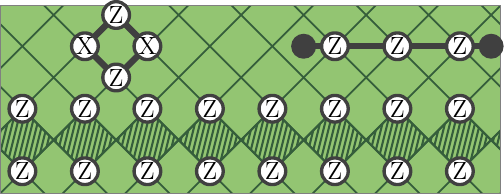}
\caption{The XZZX surface code~\cite{Bonilla-Ataides:2021} is defined with a qubit on each vertex of a square lattice. Each face of the lattice has a single stabilizer, the product of Pauli operators $XZZX$. We show a stabilizer to the top left of the figure. The code has one-dimensional symmetries with respect to an error model that is biased to introduce dephasing errors only. Indeed, the product of the hatched stabilizers commutes with all Pauli-Z errors. This gives rise to string-like errors that are constrained to align along horizontal lines under this noise bias. We show one such error to the top right of the figure. This structure is advantageous to the design of high-threshold decoding algorithms for highly-biased noise models. \label{Fig:XZZX}}
\end{figure}

A system symmetry $\Sigma$ is a subset of stabilizers such that
\begin{equation}
\left( \prod_\Sigma S \right) E |\psi\rangle = (+1)E|\psi \rangle, \quad \forall E \in \mathcal{E}. \label{Eqn:SystemSymmetry}
\end{equation}
A system symmetry implies that we can decode using minimum-weight perfect matching by pairing events on a smaller subset of stabilizer generators. This is due to the fact that we may obtain different defect parity conservation laws if we concentrate on specific error models $\mathcal{E} \subset\mathcal{P}^{\otimes n}$. This can qualitatively change the way we design decoders. We note that the definition for a materialised symmetry $\Sigma$ given earlier, see Eqn.~(\ref{Eqn:MatSymmetry}), will trivially satisfy the relationship for a system symmetry, Eqn.~(\ref{Eqn:SystemSymmetry}), for any error model. In this sense, a system symmetry represents a generalisation of the definition for a materialised symmetry proposed earlier.

Let us take the XZZX code as an example~\cite{Bonilla-Ataides:2021}. The code has a single stabilizer operator for each face of a square lattice, see Fig~\ref{Fig:XZZX}. The XZZX code is specalised to correct for dephasing noise where the error model $ \mathcal{E}^Z $ is generated by single-qubit Pauli-Z operators. The XZZX code is locally equivalent to the standard surface code via single-qubit rotations. However, unlike the surface code, under an infinitely biased-noise dephasing model that introduces Pauli-Z type errors only, $\mathcal{E}^Z$, the XZZX code demonstrates one-dimensional system symmetries. We show the stabilizer generators that multiply to give one such symmetry by the hatched faces in the figure. Given that the product of these terms commutes with Pauli-Z errors, these symmetries satisfy Eqn.~(\ref{Eqn:SystemSymmetry}).

The consequences of the one-dimensional system symmetry mean that all string-like errors from the Pauli-Z model are constrained to have a common orientation. We show one such string error with a horizontal orientation to the top right of Fig.~\ref{Fig:XZZX}. This structure means that we can decode the errors on one-dimensional lines separately~\cite{Bonilla-Ataides:2021}, as though each row is a disjoint repetition code. We therefore obtain an exceptionally high threshold in the infinitely biased limit~\cite{Bonilla-Ataides:2021}. Further to this, the XZZX code can be generalised to correct for general, finitely biased Pauli noise. A decoder for this code is obtained readily using the fact that the XZZX code is locally equivalent to the surface code, such that we can adapt the standard decoding algorithms for this purpose~\cite{Bonilla-Ataides:2021}.

\begin{figure}[b]
\includegraphics{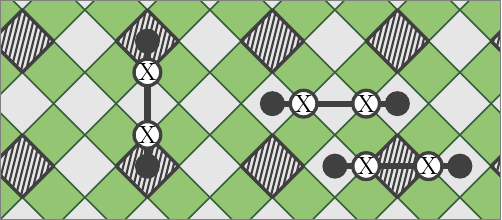}
\caption{\label{Fig:Ballistic} The ballistic noise model acting on the surface code with Pauli-X(Pauli-Z) stabilizers on the green(grey) faces. The noise model introduces strings of bit-flip errors to the surface code of a fixed length $\xi$ that are oriented either vertically or horizontally. The figure shows errors of fixed length $\xi = 2$. The product of the hatched face operators respect a defect parity conservation law, where the stabilizer operators are separated by a regular distance. We therefore obtain a system symmetry for the surface code with respect to the ballistic noise model, by taking the set of hatched faces, and translations thereof.}
\end{figure}

In practice, error models that occur in nature may typically introduce correlated errors. One might consider finding system symmetries with other error models where perhaps the Pauli error model $\mathcal{E}$ introduces correlated errors that respect a certain structure. For instance, in Ref.~\cite{Nickerson19} a correlated noise model acting on the surface code called the ballistic noise model is considered, see Fig.~\ref{Fig:Ballistic}. The noise model introduces strings of bit-flip errors where the strings are constrained along certain orientations and have a fixed length. The rigid structure of the error model is such that decoding for the ballistic noise model can be decomposed into matching on subsets of the stabilizers of the surface code with a regular separation. The stabilizer generators $S_f$ of one such symmetry are marked by hatched faces in Fig.~\ref{Fig:Ballistic}. It may be interesting to find other correlated noise models that respect system symmetries of codes, particularly if those correlated errors frequently occur due to real physical processes. Ideally, we might try to seek an optimal choice of code $\mathcal{S}$ such that a natural error model $\mathcal{E}$ gives rise to many system symmetries that can be exploited for a high-performance decoding.

\subsection{Towards a generalised matching decoder}

Having mentioned a number of examples where minimum-weight perfect-matching decoders have been obtained using symmetries of the stabilizer group, a natural question to ask is how we might choose a suitable symmetry for decoding with more general stabilizer codes. Let us finally discuss what such a generalisation might look like. We propose a construction that gives rise to a matching decoder for codes with certain properties. We present the construction by giving a systematic way of reproducing the surface-code decoder that we discussed in detail in SubSec.~\ref{SubSec:MWPM}. We point to the generic properties of which we make use to reproduce the decoder using our construction, and we discuss some questions that we should address to find a general construction for a decoding algorithm that makes use of minimum-weight perfect matching.

We require a systematic way of finding a non-independent set of stabilizer generators to form a symmetry $\Sigma$ that will be suitable for identifying high-weight errors with a matching decoder. Here we propose using the notion of cleaning~\cite{Bravyi_2009} to obtain one such set of stabilizers. The act of cleaning means to change the support of a logical operator by multiplication by stabilizer operators. We posit that the boundary operator $b$, shown in Fig.~\ref{Fig:SurfaceCode}(f), is an operator that cleans the logical operator a long way over the two-dimensional qubit array, from one boundary to another. More specifically, $b$ cleans one instantiation of the logical operator $\overline{Z}$, that is supported on the top boundary of the lattice, onto another, $\overline{Z'}$, supported at the bottom boundary. The two instantiations of logical operators are such that $\overline{Z}\cdot \overline{Z'} = b \in  \mathcal{S}$.

Let us now look at how we use the boundary operator to find a correction using a perfect-matching algorithm. To produce a correction operator $C$ that corrects the code successfully with high probability, it is sufficient to find the commutator of the error $E$ on a canonical set of logical operators, say $\overline{X}_j$ and $\overline{Z}_j$. Given this information, and an arbitrary choice of correction $C'$ that recovers any state in the code space, we can always multiply $C'$ by a logical operator $L$ such that $C = C'L$ where $C$ has the same commutator with the logical operator as $E$. It will follow that $CE$ commutes with all logical operators, therefore showing that $CE\in \mathcal{S}$, by definition. In the following discussion we note that we can determine the commutator of $E$ with each logical operator individually using matching. We therefore consider only a single logical operator, say $\overline{Z}$, and assume that we can repeat this process for all logical operators to determine a correction to recover the encoded state.

We use the stabilizer elements of $\Sigma$ to attempt to determine the commutator of $E$ with $\overline{Z}$. By definition we obtain a symmetry by taking $b$ together with the local stabilizer generators used to generate $b$. The value of $b$ determines the commutator of the error $E$ on two instances of the logical operator, $\overline{Z} \cdot \overline{Z'} $. It is our goal to use the the values of the stabilizers of the symmetry $\Sigma$ to distinguish the support of the error on $\overline{Z} $ rather than $\overline{Z'}$.

Let us now point to why we have cleaned $\overline{Z'} $ far from $\overline{Z}$. Local errors supported on $\overline{Z}$ demonstrate a very different syndrome compared with that of local errors supported on $\overline{Z'} $. In addition to changing the parity of $b$, local errors supported on $\overline{Z}$ also create defects on a faces that is close to the top boundary in order to maintain the defect parity conservation law using local stabilizer generators. In contrast, local errors supported on $\overline{Z'}$, that also change the value of $b$, will also create defects that are close to the bottom boundary.

The minimum-weight perfect-matching algorithm exploits the difference in the syndrome for local errors supported on $\overline{Z} $ and $\overline{Z'}$ to determine the commutator of $E$ with $\overline{Z} $. Specifically, each defect for stabilizers in $\Sigma$, including $b$, is assigned a vertex for the graph that is input to the perfect-matching algorithm. 
Edges connecting the input vertices are then assigned low weights if it is highly probable that the noise model caused some error event that created the given pair of defects, and higher weights otherwise.

We use the solution of the minimum-weight perfect matching to determine how many errors are supported on $\overline{Z}$ rather than $\overline{Z'}$. We find that parity of the number of defects paired via the top boundary tells us precisely the support of local errors on $\overline{Z}$. Specifically, an even number of edges paired via the top boundary indicates that $E$ commutes with $\overline{Z}$ and an odd number of edges paired via the top boundary indicates that $E$ anti commutes with $\overline{Z}$.

The decoder will fail if defects are mistakenly paired onto $\overline{Z'}$ instead of $\overline{Z}$, or vice versa. However, given that $b$ cleans $\overline{Z}$ a long distance over the lattice onto $\overline{Z}'$, it should be easy to determine which defects should be paired to $\overline{Z}$ for typical errors that occur at low-error rates, because only high-weight errors that find adversarial configurations will lead to a failure~\cite{Dennis02}.

This example demonstrates a way for how we might generalise a minimum-weight perfect-matching decoder to other codes. Let us first summarise our construction assuming an arbitrary choice of code with certain properties that we state shortly. The construction is also presented in Fig.~\ref{Fig:xCube} for the example of the X-cube model. We have proposed beginning with some instance of a logical operator, $\overline{Z}$, and then used local stabilizer generators to clean $\overline{Z}$ onto another instantiation of the logical operator $\overline{Z'}$ supported onto a distant set of qubits with operator $b \in \mathcal{S}$ . We then use the defects for local stabilizer generators $S \in \Sigma \backslash b $ for symmetry $\Sigma$ such that 
\begin{equation}
b = \prod_{\Sigma \backslash b } S,
\end{equation}
to attempt to distinguish the support of error $E$ on logical operator $\overline{Z}$ rather than $\overline{Z'}$. We make this determination using a sensible strategy for matching defects onto either the support of $\overline{Z}$ rather than $\overline{Z'}$ that should be well separated for $b$.

The two key properties of the surface code we have used to propose this construction are: (I)~that $\overline{Z}$ is well separated from $\overline{Z'}$ over the qubit array, and (II)~that we knew a suitable initial choice for $\overline{Z}$ to obtain a suitable choice of $b$. We expect that this construction will reproduce minimum-weight perfect-matching decoders with codes where we have a common set of properties. However, we do not expect codes to satisfy these properties in general. There remain several questions, whose answers may allow us to relax the assumptions we have used to obtain our systematic construction.

First of all, the construction proposed above assumes that we begin with a logical operator $\overline{Z}$. However, there are many different choices for an initial logical operator. We can make any choice $S \overline{Z}$ for all stabilizers $S \in \mathcal{S}$. In the example of the surface code, we deliberately began with a logical operator that reproduced the standard decoder~\cite{Dennis02} where $\overline{Z}$ is supported on the top boundary. However, one could imagine reproducing this strategy with a bad initial choice of logical operator such that the resulting matching algorithm has suboptimal performance. It will be interesting to find a systematic way of finding a good choice of initial logical operator if, indeed, one can always be found. It will certainly be valuable to continue looking for examples of matching decoders for LDPC codes to find ways to relax this assumption.

We have also assumed that a logical operator $\overline{Z}$ can be cleaned onto $\overline{Z'} = b \overline{Z}$ such that $\overline{Z'}$ is supported on qubits that are far away from the support of $\overline{Z}$. However, not all codes have this property. Take for example the color code with triangular boundaries~\cite{Bombin06}. In this example, all instantiations of $\overline{Z} $ and $ \overline{Z'}$ must have at least one common qubit in their support. In spite of this, matching decoders have been demonstrated, see, e.g., Refs.~\cite{Wang10graphical, Delfosse14, Kubica19, Brown20parallelized, Nixon21, Sahay22}.

Take as another related case, codes that have periodic boundary conditions, the surface code embedded on a torus being the prototypical example~\cite{Kitaev03}. In this case we obtain the symmetry that is typically used to decode this code by cleaning the logical operator $\overline{Z}$ all the way around the torus and back onto itself, such that $\overline{Z} = \overline{Z'}$, using the symmetry whereby $\prod_{f \in \text{grey}} S_f =\mathds{1}$. These two examples suggest that we may be able to design a general minimum-weight perfect-matching decoder by relaxing the assumption that $\overline{Z} $ needs to be cleaned far away from its initial support. Indeed, the reader that has followed this discussion up to this point may have noticed that at no stage has the notion of `far away' been defined precisely.

\begin{figure}
\begin{center}
\includegraphics{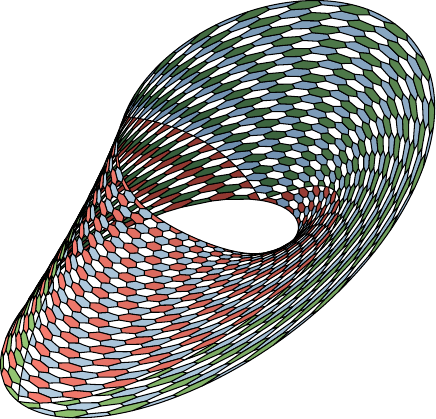}
\end{center}
\caption{The color code with three different boundary types on the sides of a triangular qubit array gives rise to a symmetry supported on a M\"obius strip, where each stabilizer generator of the code appears twice over the manifold. \label{Fig:Moebius}}
\end{figure}

Building on the last point, let us finally mention that matching decoders can be designed where the manifold on which matching is conducted does not reflect the topology of the code. In one example~\cite{Sahay22}, a symmetry is found where the local stabilizer generators of a planar color code are embedded on the surface of a M\"obius strip. We show this symmetry in Fig.~\ref{Fig:Moebius}. Interestingly, each stabilizer generator of the planar color code appears twice on the M\"obius strip. Likewise, the decoder used to correct the surface code with boundaries under a biased noise model uses a symmetry where each stabilizer appears twice~\cite{Tuckett20}. A general approach to discover the matching procedure used in these decoders may require a method more general than cleaning, that allows the duplication of stabilizers in a given symmetry, together with a way of presenting the manifold on which the symmetry is supported. A clearer understanding of how to produce and describe these symmetries may reveal new structure in general LDPC codes with non-local stabilizers~\cite{Breuckmann21}, on the path towards a general construction for a minimum-weight perfect matching decoder.

\section{Conclusion}
\label{Sec:Discussion}
Fault-tolerant quantum computing hardware will require a high-performance decoding algorithm to correct for the errors they suffer. We have reviewed some of the progress that has been made towards the development of the surface code; one of the leading architectures that is now in production to demonstrate scalable quantum computing. In particular we have reviewed many of the decoding algorithms that have been developed to support this practical fault-tolerant quantum-computing system. The minimum-weight perfect-matching decoder has been central to our discussion. This is among the most widely studied decoders that has been considered for surface-code error correction.

In the example of the surface code, the minimum-weight perfect-matching decoder makes use of a parity conservation law, which arises due to a materialised symmetry. This property is fundamental to the structure of the surface code. By presenting this perspective, we hope to find ways of generalising the minimum-weight perfect-matching decoder using similar structures in other codes. To this end, we have reviewed several examples of different matching decoders, and we have presented a systematic construction to produce a matching decoder for other code families, where these families satisfy certain properties. Using these examples we argue that we may be able to go beyond our construction by relaxing some of the assumptions we have used. In future work, it will be exciting to discover new decoders by exploiting the underlying symmetries and conservation laws of general stabilizer codes.

\section*{Acknowledgment}

The author would like to thank S. Bartlett, S. Bravyi, K. Brown, A. Cross, S. Flammia and D. Williamson for helpful and encouraging conversations.

\ifCLASSOPTIONcaptionsoff
  \newpage
\fi

\end{document}